\documentstyle[11pt,aaspp4,flushrt]{article}

\newcommand{\celr}{C_{\rm elr}}
\newcommand{\cc}{C_{\rm c}}
\newcommand{\cf}{C_{\rm f}}
\newcommand{\etal}{{et~al.}}
\newcommand{\taue}{\tau_{\rm elr}}
\newcommand{\tauc}{\tau_{\rm c}}
\newcommand{\zabs}{{z_{\rm abs}}}
\newcommand{\zem}{{z_{\rm em}}}
\newcommand{\DR}{{{f_{\rm b} \lambda_{\rm b}} \over {f_{\rm r} \lambda_{\rm r}}}}

\newcommand{\CIVdblt}{{\rm C}\kern 0.1em{\sc iv}~$\lambda\lambda 1548, 1550$}
\newcommand{\MgIIdblt}{{\rm Mg}\kern 0.1em{\sc ii}~$\lambda\lambda 2796, 2803$}
\newcommand{\NVdblt}{{\rm N}\kern 0.1em{\sc v}~$\lambda\lambda 1238, 1 242$}
\newcommand{\OVIdblt}{{\rm O}\kern 0.1em{\sc vi}~$\lambda\lambda 1031, 1037$}
\newcommand{\SiIVdblt}{{\rm Si}\kern 0.1em{\sc iv}~$\lambda\lambda1393, 1402$}
\newcommand{\AlII}{\hbox{{\rm Al}\kern 0.1em{\sc ii}}}
\newcommand{\AlIII}{\hbox{\rm Al}\kern 0.1em{\sc iii}}
\newcommand{\CaII}{\hbox{{\rm Ca}\kern 0.1em{\sc ii}}}
\newcommand{\CII}{\hbox{{\rm C}\kern 0.1em{\sc ii}}}
\newcommand{\CIIe}{\hbox{{\rm C$^{\ast}$}\kern 0.1em{\sc ii}}}
\newcommand{\CIII}{\hbox{{\rm C}\kern 0.1em{\sc iii}}}
\newcommand{\CIV}{\hbox{{\rm C}\kern 0.1em{\sc iv}}}
\newcommand{\CV}{\hbox{{\rm C}\kern 0.1em{\sc v}}}
\newcommand{\NHI}{\hbox{$N$({\rm H}\kern 0.1em{\sc i})}} 
\newcommand{\HI}{\hbox{{\rm H}\kern 0.1em{\sc i}}}
\newcommand{\HII}{\hbox{{\rm H}\kern 0.1em{\sc ii}}}
\newcommand{\Lya}{\hbox{{\rm Ly}\kern 0.1em$\alpha$}}
\newcommand{\Lyb}{\hbox{{\rm Ly}\kern 0.1em$\beta$}}
\newcommand{\Lyg}{\hbox{{\rm Ly}\kern 0.1em$\gamma$}}
\newcommand{\Lyfive}{\hbox{{\rm Ly}\kern 0.1em$5$}}
\newcommand{\Lysix}{\hbox{{\rm Ly}\kern 0.1em$6$}}
\newcommand{\Lyseven}{\hbox{{\rm Ly}\kern 0.1em$7$}}
\newcommand{\Lyeight}{\hbox{{\rm Ly}\kern 0.1em$8$}}
\newcommand{\Lynine}{\hbox{{\rm Ly}\kern 0.1em$9$}}
\newcommand{\Lyten}{\hbox{{\rm Ly}\kern 0.1em$10$}}
\newcommand{\HeI}{\hbox{{\rm He}\kern 0.1em{\sc i}}}
\newcommand{\HeII}{\hbox{{\rm He}\kern 0.1em{\sc ii}}}
\newcommand{\FeI}{\hbox{{\rm Fe}\kern 0.1em{\sc i}}}
\newcommand{\FeII}{\hbox{{\rm Fe}\kern 0.1em{\sc ii}}}
\newcommand{\FeIII}{\hbox{{\rm Fe}\kern 0.1em{\sc iii}}}
\newcommand{\MnII}{\hbox{{\rm Mn}\kern 0.1em{\sc ii}}}
\newcommand{\MgI}{\hbox{{\rm Mg}\kern 0.1em{\sc i}}}
\newcommand{\MgII}{\hbox{{\rm Mg}\kern 0.1em{\sc ii}}}
\newcommand{\MgIII}{\hbox{{\rm Mg}\kern 0.1em{\sc iii}}}
\newcommand{\NV}{\hbox{{\rm N}\kern 0.1em{\sc v}}}
\newcommand{\NII}{\hbox{{\rm N}\kern 0.1em{\sc ii}}}
\newcommand{\NIII}{\hbox{{\rm N}\kern 0.1em{\sc iii}}}
\newcommand{\OVIII}{\hbox{{\rm O}\kern 0.1em{\sc viii}}}
\newcommand{\OVII}{\hbox{{\rm O}\kern 0.1em{\sc vii}}}
\newcommand{\OVI}{\hbox{{\rm O}\kern 0.1em{\sc vi}}}
\newcommand{\OII}{\hbox{[{\rm O}\kern 0.1em{\sc ii}]}}
\newcommand{\SiI}{\hbox{{\rm Si}\kern 0.1em{\sc i}}}
\newcommand{\SiII}{\hbox{{\rm Si}\kern 0.1em{\sc ii}}}
\newcommand{\SiIII}{\hbox{{\rm Si}\kern 0.1em{\sc iii}}}
\newcommand{\SiIV}{\hbox{{\rm Si}\kern 0.1em{\sc iv}}}
\newcommand{\SII}{\hbox{{\rm S}\kern 0.1em{\sc ii}}}
\newcommand{\SIII}{\hbox{{\rm S}\kern 0.1em{\sc iii}}}
\newcommand{\SIV}{\hbox{{\rm S}\kern 0.1em{\sc iv}}}
\newcommand{\NaI}{\hbox{{\rm Na}\kern 0.1em{\sc i}}}
\newcommand{\kms}{\hbox{km~s$^{-1}$}}
\tighten
\begin{document}

\received{1 December 1998}

\slugcomment{\it Submitted to {\it The Astronomical Journal}}
\lefthead{Ganguly \etal}
\righthead{Partial Coverage}


\title{\Large\bf Intrinsic Narrow Absorption Lines in HIRES/Keck Spectra 
of a Sample of Six Quasars\altaffilmark{1}}
\pagestyle{empty}

\author{Rajib~Ganguly, Michael~Eracleous, Jane~C.~Charlton 
\altaffilmark{2}, and Christopher~W.~Churchill \altaffilmark{3}}
\medskip
\affil{\normalsize\rm Department of Astronomy and Astrophysics \\
       The Pennsylvania State University,
       University Park, PA 16802 \\
       e-mail: {\tt ganguly, mce, charlton, cwc@astro.psu.edu}}

\begin{center}
Submitted to {\it The Astronomical Journal}
\end{center}

\altaffiltext{1}{Based in part on observations obtained at the
W.~M. Keck Observatory, which is jointly operated by the University of
California and the California Institute of Technology.}

\altaffiltext{2}{Center for Gravitational Physics and Geometry,
                 The Pennsylvania State University}

\altaffiltext{3}{Visiting Astronomer, The W.~M.~Keck Observatory}

\pagestyle{empty}
\begin{abstract}

In the course of a large survey of intervening {\MgII} absorbers in
the spectra of quasars, the {\CIV} emission lines of six of the target
objects (three radio--loud and three radio--quiet) were observed
serendipitously. In four of these six quasars, we detected
``associated'' narrow absorption lines with velocities within
5000~{\kms} of the quasar emission--line redshift (three of these four
quasars are radio--quiet while the other is radio--loud). As a result of
the original target selection, the small sample of six quasars is
unbiased towards finding associated absorption lines.
In three of these four cases, the absorption line optical--depth
ratios deviate from the prediction based on atomic physics, suggesting
that the background photon source(s) are only partially covered by the
absorbing medium and, by extension, that the absorption lines are
intrinsic to the quasar. We have used the method of Barlow \& Sargent
to determine the effective coverage fraction of background source(s)
and we have extended it to constrain the coverage fraction of the
continuum and broad--emission line sources separately. We have also
applied this refined method to the narrow intrinsic absorption lines
in three additional quasars for which the necessary data were
available from the literature. We find that in two objects from our
sample, the {\it continuum} source must be partially covered
regardless of the covering factor of the emission--line source. We
discuss these results in the context of the properties of absorption
lines observed in different types of active galaxies and related
outflow phenomena. We cannot distinguish between possible mechanisms
for the origin of the partial coverage signature although we do
consider possible observational tests.  Finally, we speculate on how
the gas responsible for the narrow lines may be related to the
accretion disk wind that may be responsible for the broad absorption
lines observed in some quasars.

\end{abstract}

\keywords{quasars: absorption lines --- quasars:  individual
(Q~$0450-132$, Q~$1213-003$, PG~$1222+228$, PG~$1329+412$)}

\section{Introduction}
\pagestyle{myheadings}
\markboth{\small Ganguly et al. \hfill Intrinsic Quasar Absorbers~~}
         {\small Ganguly et al. \hfill Intrinsic Quasar Absorbers~~}

Absorption by an ionized medium close to the central engine is now
recognized as a fairly common feature in the spectra of active
galactic nuclei (hereafter AGNs). In Seyfert~1 galaxies, the signature
of the ionized gas takes the form of narrow UV resonance absorption
lines from highly--ionized species, such as {\NV} and {\CIV} (Crenshaw,
Maran, \& Mushotzky 1998\nocite{cmm98}) as well as soft X--ray
absorption edges from highly ionized species such as {\OVII} and
{\OVIII} (\cite{reyn97}; \cite{geo98}).  Similar UV absorption
lines\footnote{Here, we consider only absorption lines which are close
to the redshift of the quasar, i.e.  {$\zabs\approx\zem$}, not lines
at considerably lower redshifts than that of the quasar.} are also
observed in quasars (e.g. \cite{foltz86}; \cite{and87}; Young,
Sargent, \& Boksenberg 1982\nocite{ysb82}; Sargent, Boksenberg, \&
Steidel 1988\nocite{sbs88}; \cite{ss91}).  Quasars, however, are
perhaps better known for their {\it broad} absorption lines, whose
widths often reach {50,000~\kms} (e.g., \cite{turn88};
\cite{wey91}). The high ionization state of the absorbers and the
often blueshifted absorption lines (relative to the broad emission
lines) suggest that the absorber is a fairly tenuous medium outflowing
from the AGN, most likely accelerated by radiation pressure
(\cite{arav95}). The absorber could plausibly be associated with an
accretion--disk wind (e.g.,
\cite{mur95}; \cite{kk95}) analogous to those observed 
in many cataclysmic variables (e.g., \cite{drew90}; \cite{mauche94}).
This can only be a broad analogy, however, since a number of
observational clues suggest that the picture in AGNs is neither as
simple nor as ``clean'' as in cataclysmic variables. For example, the
fact that absorption line depths often exceed either the net continuum
level or the net emission--line level on which they are superposed
suggests that the absorber does not lie between the broad
emission--line region (BELR) and the more compact continuum source
since it intercepts both continuum and emission--line photons. That is,
it lies either within the BELR or away from both regions.

The relation between the absorption--line region and the ubiquitous AGN
emission--line regions is unknown, although it has been suggested that
they may represent different phases or layers of the same medium
(Shields, Ferland, \& Peterson 1995\nocite{shields95};
\cite{ham98}). From an observational perspective, it is important to
determine the geometry of the absorbing gas and its physical
conditions because these constitute useful constraints on theoretical
models of the ionization and acceleration of the outflow.  Moreover,
it is also interesting to ask whether there are any trends in the
properties of the absorption lines with AGN subclass since such trends
serve as constraints on scenarios of fundamental differences between
such subclasses (e.g., the radio--loud/radio--quiet AGN dichotomy).  It
is, therefore, encouraging that some progress has been made in recent
years in answering the above questions. First, observations of large
samples of quasars indicate that {\it broad} absorption lines (BALs)
are found preferentially among radio--quiet quasars (\cite{stocke92}),
while the preferred hosts of associated
\footnote{``Associated'' absorption lines, according to the definition of
Anderson \etal~(1987\nocite{and87}), are
those with velocities within {5000~\kms} of the quasar rest frame. In
principle, associated absorption lines need not be intrinsic to the
quasar.}
{\it narrow} absorption lines (NALs) are radio--loud quasars
(\cite{foltz86}; \cite{and87}).  These trends are by no means
well--established and have been called into question by the discovery
of radio--loud BAL quasars (\cite{beck97}; \cite{broth98}) and
radio--quiet intrinsic NAL quasars (Hamann, Barlow, \& Junkkarinen
1997a\nocite{ham97a}; \cite{ham97b}). 
We note, however, that the Foltz \etal~(1986\nocite{foltz86})
and Anderson \etal~(1987) results discuss the preference of
{\it strong} associated NALs for radio--loud quasars, while also
demonstrating that weak associated NALs occur in both radio--quiet
and radio--loud quasars.
Second, observations of
selected NAL quasars at high spectral resolution have shown that the
absorbers only partially cover the background BELR and/or continuum
source (\cite{bs97}; \cite{ham97a}). In another quasar, the NALs were
found to vary on a time scale of only a few months
(\cite{ham97b}). These observations demonstrate that the NALs in these
objects are intrinsic and also provide useful constraints on their
geometry.

In this paper, we present five associated NALs along the lines of
sight toward six quasars observed at high spectral resolution. These
were discovered serendipitously in a survey of {\it intervening} {\MgII}
absorbers ({$\zabs \ll \zem$}; \cite{cwcthesis}). In three of the five
cases, the NAL systems exhibit the partial coverage signature of
absorbing gas {\it intrinsic} to the quasar. In the process of
analyzing the data, we have refined and expanded the method of Barlow
{\&} Sargent~(1997\nocite{bs97}) to treat the coverage fractions of
the continuum source and the BELR separately. Although it is, in
principle, impossible to determine the two coverage fractions
independently using an absorption doublet, it is possible to place
interesting constraints on them. In particular, we find that in two of
the quasars in this small collection, the continuum source must be
partially covered by the absorber.

In \S2 of this paper, we describe the selection of the original quasar
sample, the observations and data reduction, and the properties of the
associated NALs that we detected and other relevant properties of
their hosts. In \S3 and \S4, we show that, in three of these systems,
the relative strengths of the absorption lines provide evidence for
partial coverage of the background source by the absorber. In \S4, we
determine the effective coverage fractions by applying the method of
Barlow \& Sargent (1997\nocite{bs97}). We also develop and apply a
refined version of this method in which we treat the coverage fraction
of the continuum source and broad BELR separately. In \S5, we discuss
the implications of the observational results, possible partial
coverage mechanisms, and the general picture of intrinsic absorbers in
different types of active galaxies. In \S6, we summarize our findings
and present our final conclusions.


\section{Sample of Objects, Observations, and Data}

The associated NAL systems reported here were found in the course of a
large statistical study of intervening {\MgII} absorbers during which
spectra of a sample of 25 quasars were obtained (\cite{cwcthesis}). In
six objects from this sample, the spectra included the broad {\CIV}
emission line and revealed the presence of five associated NAL systems
within {$5000$~\kms} of the emission--line redshift.  Another system at
$\Delta v = -13,800~{\kms}$ toward Q~$0450-132$ could be
``associated'' (see Petitjean, Rauch, \& Carswell 1994\nocite{pet94}),
but it does not satisfy the traditional velocity criterion. These six
quasars are listed in Table~\ref{tab:qso} along with the some of their
basic properties. The emission redshifts and emission line fluxes for
the quasars were taken from Sargent \etal~(1988\nocite{sbs88}) and
from Steidel \& Sargent (1991\nocite{ss91}). Because of the way the
original sample was selected, the final collection of six quasars
whose {\CIV} emission lines were observed is unbiased towards the
presence of associated {\CIV} NALs. In two of the six objects the
spectra include additional absorption lines, namely {\SiIV} and {\NV}
in Q~$0450-132$ and {\SiIV} in Q~$1213-003$.

Since the radio properties of these quasars are important to our later
discussion, we have compiled information about their radio properties
based on reports in the literature and included it in
Table~\ref{tab:qso} along with the appropriate references. The observed
5~GHz radio flux was converted to the rest frame of the source using
the measured spectral index $\alpha$, if available, or assuming
$\alpha=0.5$ otherwise (where $f_{\nu}\propto \nu^{-\alpha}$). The
radio power was computed assuming a Hubble constant of 50~${\rm
km~s^{-1}~Mpc^{-1}}$, and a deceleration parameter of ${1\over 2}$,
following Kellermann \etal~(1994\nocite{kel94}). The rest--frame
4400~{\AA} optical flux was calculated from the observed $V$ magnitude
assuming a flat optical--UV spectrum (i.e., $\alpha=0$;
\cite{elvis94}). To classify an object as radio--loud or radio--quiet, we
adopted the criteria of Kellermann \etal~(1994\nocite{kel94})
according to which radio--loud quasars have either a 5~GHz power of
$P_{\rm\, 5~GHz} > 10^{26}~{\rm W~Hz}^{-1}$ or a 5~GHz--to--4400~{\AA}
flux density ratio of $R>10$.  According to these criteria, three of
the six objects are radio--quiet and three are radio loud.

The observations were carried out with the HIRES spectrometer
(\cite{vogt94}) on the Keck~I telescope. The spectra have a resolution
of 6.6~{\kms} with a sampling rate of 3 pixels per resolution
element.  They were reduced with the IRAF\footnote{IRAF is distributed
by the National Optical Astronomy Observatories, which are operated by
AURA, Inc., under contract to the NSF.} {\tt apextract\/} package for
echelle data.  The detailed steps for the reduction are outlined in
Churchill (1997\nocite{cwcthesis}).

The details of the observations are listed in Table~\ref{tab:obs}
which also includes the parameters of the associated NAL systems that
we detected along with the lowest measurable equivalent widths
($5\sigma$ limits) in the corresponding spectrum. In this table, we give
the mean velocity of each absorption--line system relative to the peak
of the broad emission line.  NALs were found at both negative
(blueshifted) and positive (redshifted) velocities relative to the
{\CIV} broad emission--line peak. A positive velocity, such as seen in
{PG~$1222+228$} and {PG~$1329+412$}, need not imply infall toward the
quasar since the true velocity of the quasar is more accurately
measured by narrow forbidden lines than by the broad emission line
peaks. For example, in a picture in which the NAL gas is outflowing from the
central engine, the NAL redshift would imply a smaller outflow
velocity of the absorbers relative to the broad emission--line gas.

\begin{figure}[th]
\figurenum{2}
\plotone{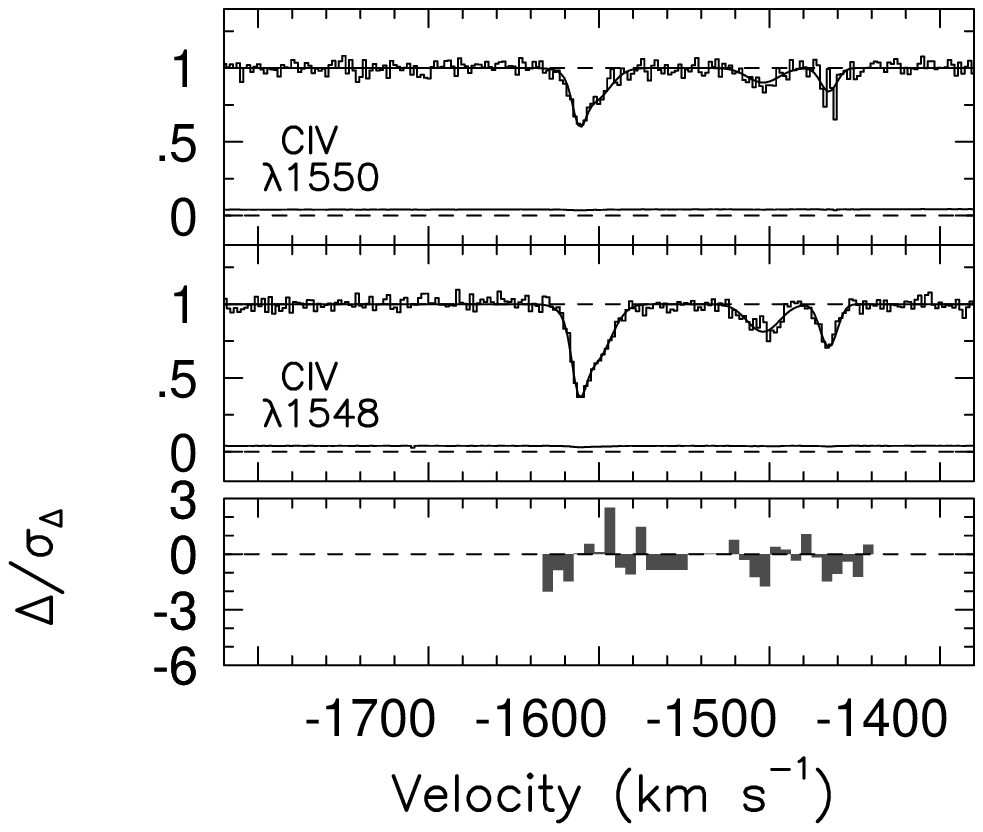}
\vglue -0.9in
\protect\caption{Same as Figure~1, but for the {$\Delta v=-1566$~\kms} 
system toward {Q~$1213-003$}.
\label{fig:q1213f1}}
\end{figure}

\begin{figure}[th]
\figurenum{3}
\plotone{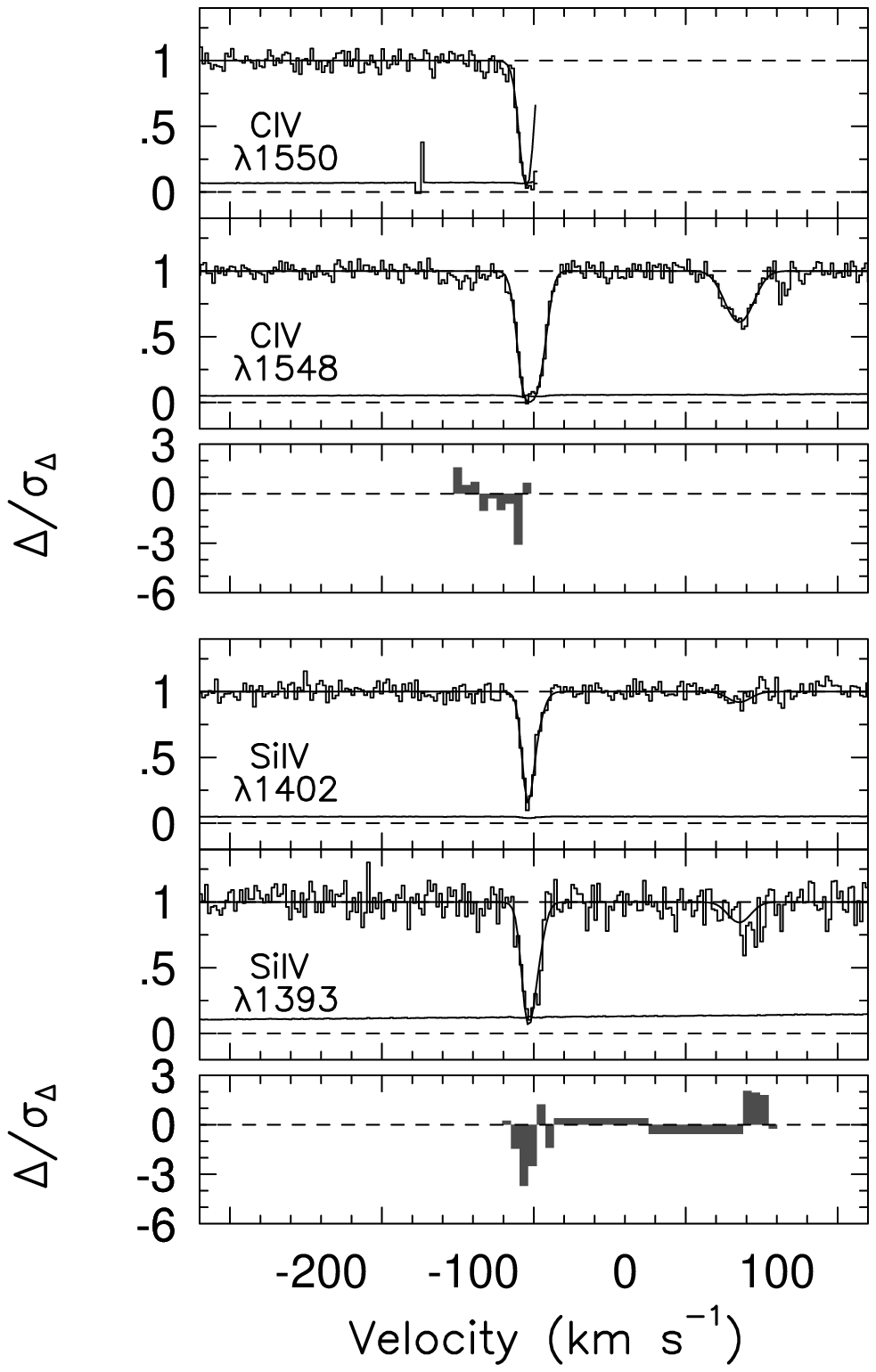}
\vglue -0.9in
\protect\caption{Same as Figure~1, but for the {$\Delta v=-60$~\kms} system toward
{Q~$1213-003$}.
\label{fig:q1213f2}}
\end{figure}

Figures \ref{fig:q0450f1}--\ref{fig:q1329f1} present the velocity
aligned absorption profiles of the {\CIV} doublets and of other
detected transitions ({\SiIV} and {\NV}) of the five associated
systems that were covered in the HIRES/Keck~I spectra.  The
{\CIV}$\lambda$1550 profile of the {$\Delta v=-60$~\kms} system of
{Q~$1213-003$} (Figure~\ref{fig:q1213f2}) is truncated near 0~{\kms}
because is falls off the edge of the CCD. In fact, because of the
redshift of {Q~$1213-003$} its {\CIV} emission line is in the
observed wavelength range above 5100~\AA, where the echelle orders are
separated by gaps. As a result, there are several gaps in the coverage
of the broad emission--line profile. None of the other objects suffer
from this problem, however.

The kinematic structure of the NAL profiles is varied and has multiple
components. It is not distinguishable from the structure found in
intervening absorbers.  However, the strong {\NV} absorption in
{Q~$0450-132$} is indicative of the typically high metallicities of
associated systems (\cite{pet94}) in contrast to the lower
metallicities of intervening systems.  In \S3 and \S4, we demonstrate
that in three of the five systems there is evidence that the absorbing
clouds only partially cover the source, which then requires them to be
in relatively close proximity to the source. It should be noted that
the {$\Delta v=-60$~\kms} system of {Q~$1213-003$} has been classified
as intrinsic by Sowinski, Schmidt, \& Hines (1997\nocite{sow97}), but
we have not been able to find the relevant report in the literature.


\section{Identification of Anomalous Optical Depths}

\begin{figure}[th]
\figurenum{1}
\plotone{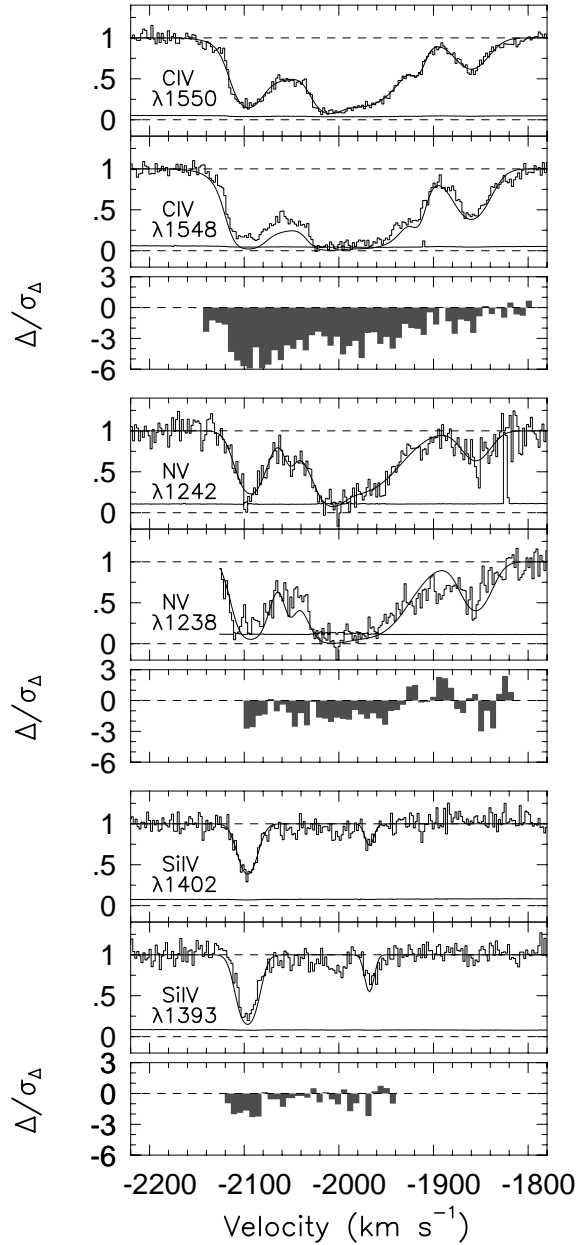}
\vglue -0.9in
\protect\caption{Velocity--aligned normalized flux profiles of detected transitions
in the {$\Delta v=-2011$~\kms} NAL system toward {Q~$0450-132$}. Model
fits to the weaker doublet members (panels 1, 4, and 7 from the top)
are shown as solid curves superimposed on the spectra. The optical
depth of each model component is scaled according to atomic physics
and overplotted on the stronger transitions (panels 2, 5, and 8).
Panels 3, 6, and 9 show the apparent deviations of the optical depths
from the predicted value averaged over a resolution element and scaled
by the error bar.  The {\CIV} doublet in this particular system clearly
shows inconsistencies in the optical scaling and is indicative of
partial coverage.
\label{fig:q0450f1}}
\end{figure}

\begin{figure}[th]
\figurenum{5}
\plotone{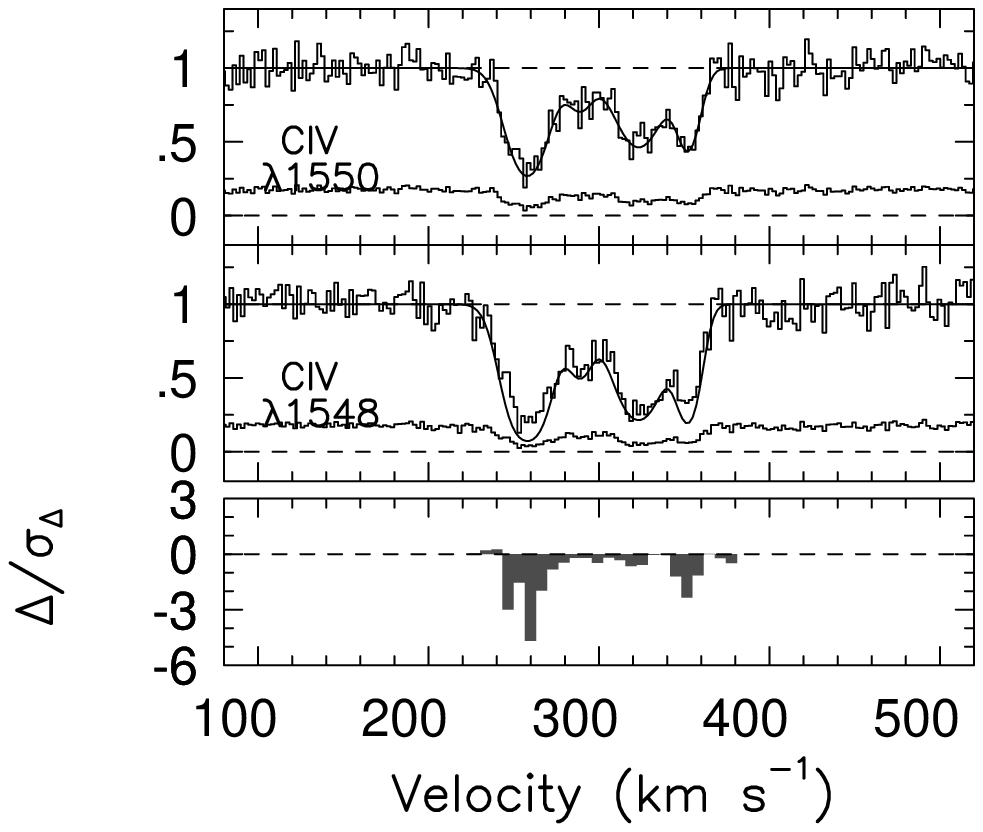}
\vglue -0.9in
\protect\caption{Same as Figure~1, but for the {$\Delta v=+314$~\kms} system toward
{PG~$1329+412$}. The obvious discrepancy between the atomic physics
prediction and the data in this particular case is indicative of
partial coverage.
\label{fig:q1329f1}}
\end{figure}

In the three of the quasars displaying NALs we find that the optical
depth ratio of the two members of the {\CIV} doublet did not have the
value expected from atomic physics, namely {${{\tau_1} \over {\tau_2}}
={{f_1 \lambda_1} \over {f_2 \lambda_2}}$} (\cite{sav91}; where $f_1$
and $f_2$ are the oscillator strengths and $\lambda_1$ and $\lambda_2$
are the rest wavelengths of the doublet members). To demonstrate this
discrepancy, we have fitted the weaker member of each doublet with a
combination of Voigt profiles, scaled the model according to what
atomic physics dictates for the stronger member of the same doublet,
and compared it to the data. The results of this exercise are
illustrated in Figures~\ref{fig:q0450f1}--\ref{fig:q1329f1}, where we
show the models as solid lines superposed on the data. The bottom
panel in each set shows the deviation of the data from the predicted
profile of the strongest member of each doublet in each resolution
element, scaled by the error bar.  We see three obviously discrepant
cases: the $\Delta v=-2011$~{\kms} system toward Q~$0450-132$
(Figure~\ref{fig:q0450f1}), the $\Delta v=+1482$~{\kms} system toward
PG~$1222+228$ (Figure~\ref{fig:q1222f1}), and the $\Delta
v=+314$~{\kms} system toward PG~$1329+412$ (Figure~\ref{fig:q1329f1}).
Neither of the associated systems toward Q~$1213-003$ appear to
deviate from the atomic physics prediction.

We consider five explanations for the apparent violation of the
optical depth scalings in these three associated systems: (1) an
instrumental/reduction effect such as scattered light which mimics a
source function (e.g., \cite{cmm98}) (2) blends with other transitions
at other redshifts; (3) a source function that fills in the line cores
(Wampler, Chugai, \& Petitjean 1995\nocite{wam0059}); (4) incomplete
occultation of the continuum and/or emission--line by the absorbers
(Wampler, Bergeron, \& Petitjean 1993\nocite{wam2116};
\cite{wam0059}; Hamann \etal~1997a,b\nocite{ham97a}\nocite{ham97b};
\cite{bs97}) and (5) scattering of background photons back into the light path.
The first
is an unlikely resolution to the problem since the optical depth
scaling is obeyed by all 64 intervening absorption--line systems
(including transitions from {\CIV}, {\MgII}, and {\FeII}) observed
with the same instrumental setup and reduced in the same manner
(\cite{weakmgII}).  In addition, we have
searched the spectra of these quasars for an extensive set of
transitions at the redshifts of all other known systems, and found
that none of them affect the profiles studied here.  This rules out
the second possibility. Some forms of the source function explanation
could be viable while others are not; we consider these further in
\S5.2. The fourth and fifth explanations have identical signatures, 
since we can think of scattering as a form of ``effective partial
coverage''.  Intuitively, the partial coverage effect will cause
models that assume full coverage to over--predict the strength of the
stronger member of a doublet, relative to the weaker.  Physically, the
intensity observed in the core of an absorption profile is the
composite of two types of light paths: (1) light that is not occulted
by the absorbing clouds or is scattered back into the light path, and
(2) the light that successfully passes through the absorbers.  The
unocculted fraction (in normalized units) is nearly the same for both
members of the doublet provided the emission line is not too steep
over the wavelength range covered by the absorption (see Appendix).
This results in a larger increase in the flux, relative to the full
coverage expectation, detected in the deeper blue member of the
doublet relative to the red member. The most straightforward example
is a situation in which the absorbers are optically thick such that
the unattenuated flux is negligible at either member of the doublet.
With partial coverage, one still observes a significant flux which is
equal in the two members.


\section{Partial Coverage}

\subsection{Effective Coverage Fraction $\cf$}

To pursue the partial coverage interpretation further we have computed
the effective coverage fraction following Barlow \& Sargent (1997) and
Hamann {\etal} (1997a). The effective coverage fraction is computed by
considering the fraction of all photons of a given wavelength that do
not pass through the absorbing gas. This calculation assumes a single,
{\it extended} continuum and emission--line source and a single
effective optical depth appropriate to the absorbing clouds at the
corresponding velocity. If {$\tau$} is the effective optical depth of
an absorbing cloud that occults a fraction {$\cf$} of the source,
the observed residual intensity, $R$, in normalized
units is:
\begin{equation}
R(\lambda) = [1-\cf(\lambda)] + \cf(\lambda) e^{-\tau(\lambda)} \quad .
\end{equation}
The first term on the right--hand side represents the photons traveling
along unocculted lines of sight while the second term arises from
photons that survive absorption due to the finite optical depth of the
cloud.

\begin{figure}[th]
\figurenum{6}
\plotone{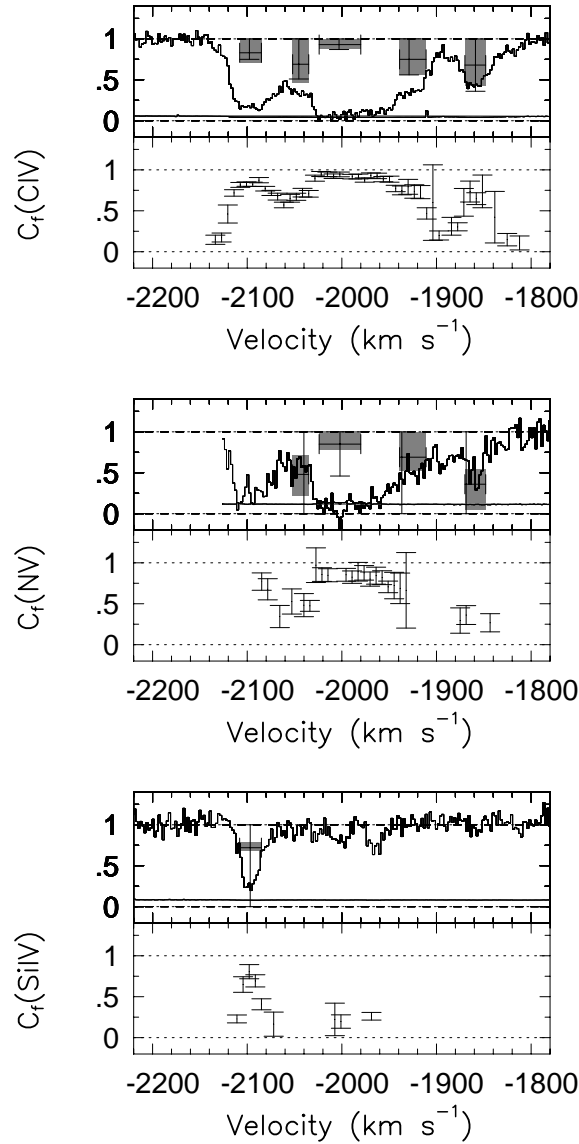}
\vglue -0.7in
\protect\caption{For each detected species in the {Q~$0450-132$} intrinsic
NAL system,
the effective coverage fraction profile, {$\cf(v)$}, with {$1\sigma$}
error bars is shown in the bottom windows of each panel.  In the top
windows, the line profiles are shown for reference along with a
graphical representation of the emission--line region and continuum
source coverage fractions averaged over 20--40~\kms--wide velocity
bins.  The level of the horizontal bars represents the weighted--mean
{$\cf$} computed over a velocity range given by its width. The
vertical bar intersects the horizontal bar at the weighted--mean
velocity and represents the allowed coverage fraction range for the
emission line region. The allowed coverage fraction ranges for the
continuum are given by the vertical extent of the shaded boxes. See
\S4.2.
\label{fig:q0450f2}}
\end{figure}

\begin{figure}[th]
\figurenum{7}
\plotone{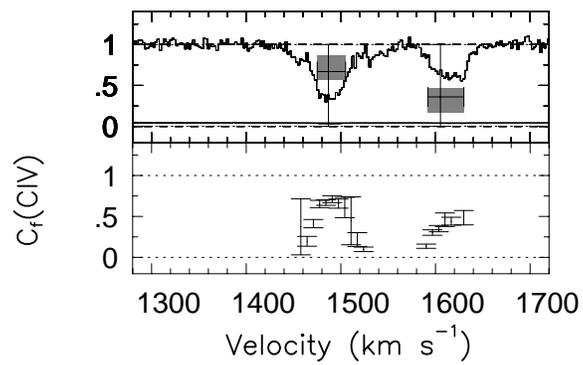}
\vglue -0.7in
\protect\caption{Same as Figure~6, but for the {PG~$1222+228$} intrinsic system.
\label{fig:q1222f2}}
\end{figure}

\begin{figure}[th]
\figurenum{8}
\plotone{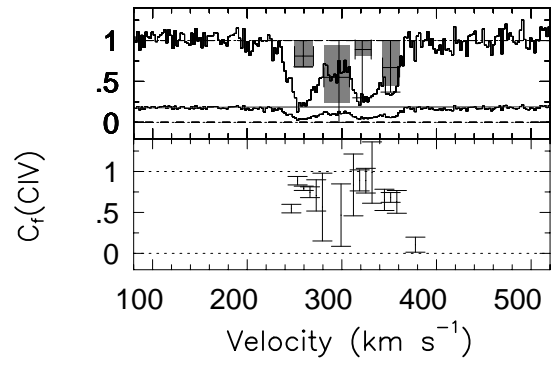}
\vglue -0.7in
\protect\caption{Same as Figure~6, but for the {PG~$1329+412$} intrinsic system.
\label{fig:q1329f2}}
\end{figure}

For two multiplet transitions, the conjunction of the two residual
intensities with the optical depth scaling yields the coverage
fraction through the solution to:
\begin{equation}
\left [{{R_{\rm r}-1+\cf} \over {\cf}}  \right ]^{{f_{\rm b} \lambda_{\rm b}} 
\over
{f_{\rm r} \lambda_{\rm r}}} = {{R_{\rm b}-1+\cf} \over {\cf}} \quad ,
\end{equation}
where the subscripts ``r'' and ``b'' refer to the properties of the
redder and bluer transitions, respectively.  For the resonant UV
doublets, such as {\CIVdblt}, {$\DR \approx 2$} allowing an analytic
solution:
\begin{equation}
\cf(v) = {\left[R_{\rm r}(v)-1\right]^2 \over {R_{\rm b}(v)-2R_{\rm r}(v)+1}} 
\quad ,
\end{equation}
where we have converted the wavelength dependence of the coverage
fraction to a velocity dependence. In the application of this method
to the data, the calculation of {$\cf$} was performed in each
resolution element along each of the resonant doublet profiles for the
three associated systems with anomalous optical depth ratios.  The
results are illustrated as points with {$1\sigma$} error bars in the
bottom window of each panel of
Figures~\ref{fig:q0450f2}--\ref{fig:q1329f2}.  Points are only plotted
if: (1) the stronger transition is detected, using the aperture method
(Lanzetta, Turnshek, \& Wolfe 1987\nocite{ltw87}), at the $3\sigma$
level; (2) the derived coverage fraction is physical ($0 < {\cf} <
1$); (3) the $1\sigma$ error in {$\cf$}, $\sigma_{\cf} < 0.5$; and (4)
the fractional uncertainty in ${\cf}$ is less than unity.

\begin{figure}[th]
\figurenum{9}
\plotfiddle{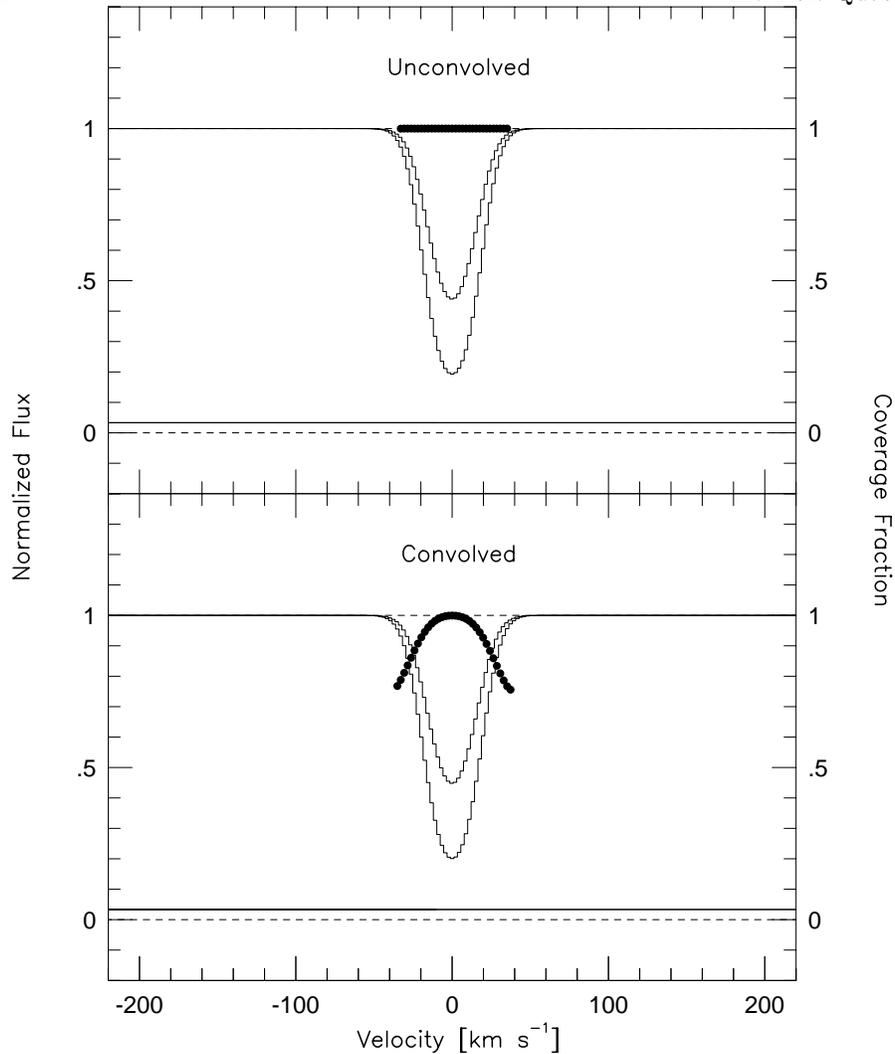}{5.0in}{0}{60}{60}{-200}{00}
\protect\caption{Convolution with the instrument's line spread function affects the
computation of the effective coverage fraction profile, {$\cf(v)$},
more in the wings of components than in their cores. The two windows
show simulated flux profiles of the {$\CIV$} doublet at infinite
signal--to--noise ratio, but finite sampling, along with the coverage
fraction derived from them (plotted as a series of circles). The top
window shows the theoretical profiles before convolution with the
instrumental line--spread function. The coverage fraction derived from
them is unity at all velocities.  However, when convolved with the
HIRES/Keck~I instrumental profile ($R\sim6.6$~\kms) (bottom window),
the wings are smeared out and the original line ratio in the wings is
distorted. As a result the coverage fraction measured in the wings is
falsely determined to be less than unity. For more details see \S4.1
of the text.
\label{fig:inseff}}
\end{figure}

A notable trend is that the coverage fraction always drops toward the
wings of lines. This is an effect of the instrument's line spread
function which tends to ``wash out'' the wings and artificially cause
anomalous optical depth ratios. We have explored this effect by
simulating observed spectra with normal optical depth ratios and
analyzing them in the same way as the data. As shown in
Figure~\ref{fig:inseff}, synthetic spectra which are not convolved
with the line spread function exhibit no variation in the coverage
fraction across the single component Voigt profile. When convolved
with the instrumental resolution function, even a profile with
${\cf}=1$ will show ${\cf}<1$ in its wings due to this convolution
effect. The signal--to--noise ratio of the observed spectra was not high
enough to merit an attempt to deconvolve the line spread function.
Therefore it is only possible to interpret the derived ${\cf}$ values
in the cores of well--resolved lines.

There is evidence for partial coverage (${\cf} < 1$) over a
significant range of velocity for the three systems. The values of
${\cf}$, averaged over the velocity regions defined by the {\it
apparent} location of kinematic components in the profile, are given
in Table~\ref{tab:covfac2}.  The weighted mean values of ${\cf}$,
averaged over all resolution elements in the regions, are also shown
graphically in the top windows of
Figures~\ref{fig:q0450f2}--\ref{fig:q1329f2}, where they are
represented by the level of the horizontal bars. The width of the
horizontal bars depicts the velocity bin over which the average value
of $\cf$ was computed. (Note that the vertical bar does not represent
an error bar but rather it indicates the coverage fraction derived
separately for the BELR, as discussed in the next section.)

We have also found that the effective coverage fraction varies with
velocity component, as follows.  In the case of Q~$0450-132$
(Figure~\ref{fig:q0450f2}), the spectrum includes absorption lines
from three species, {\CIV}, {\SiIV}, and {\NV}.  The coverage fraction
was derived in five separate regions of the {\CIV} profile and there
is evidence that it varies from component to component.  The {\NV}
data are noisy, but consistent with the coverage fraction derived for
{\CIV} in regions of overlap.  The {$\Delta v=-2100~{\rm km~s^{-1}}$}
component that shows strong {\SiIV} absorption (presumably originating
from a lower--ionization region of the absorber) also shows partial
coverage consistent with the corresponding component of {\CIV}.  There
are two distinct components in the PG~$1222+228$ intrinsic {\CIV}
absorption profile (Figure~\ref{fig:q1222f2}), both of which yield a
small coverage fraction: the {$\Delta v=+1469$~\kms} component gives
$\cf\approx 0.7$ while the {$\Delta v=+1587$~\kms} component gives
a coverage fraction of about $\cf\approx 0.4$.  The effective coverage
fraction was also determined in four separate regions of the intrinsic
{\CIV} profile of PG~$1329+412$ (Figure~\ref{fig:q1329f2}); in two of
which we found ${\cf}<1$.

\subsection{Partial Coverage of Continuum and Emission Line Sources}

\begin{figure}[th]
\figurenum{10}
\plotfiddle{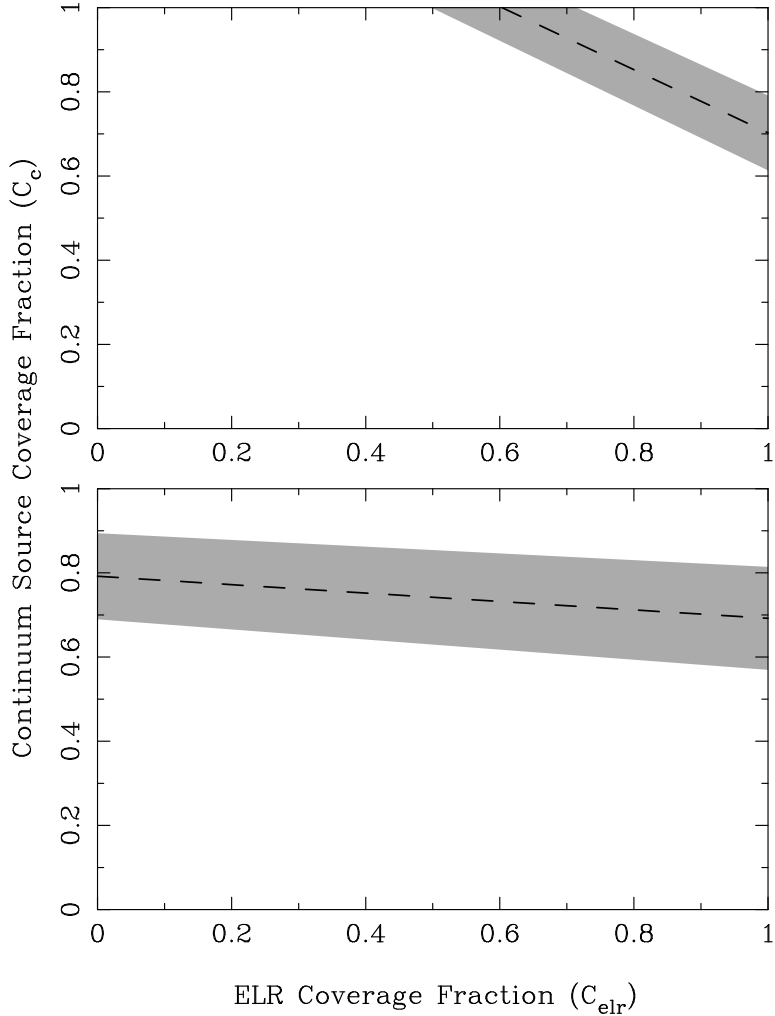}{5.0in}{0}{60}{60}{-200}{00}
\protect\caption{The $\cc$--$\celr$ parameter plane showing the solutions
allowed by
equation~(5). Given the measured values of the effective coverage
fraction, $\cf$, and the local line--to--continuum flux ratio, $W$,
equation~(5) implies a relation between $\cc$ and $\celr$. Two
examples of such a relation are plotted in the two panels of this
figure; they are described by $\cf=0.83\pm0.02,\; W=0.75\pm0.02$ and
$\cf=0.72\pm0.08,\; W=0.10\pm0.02$, respectively. The thick dashed
lines show the relation implied by the nominal values of the measured
parameters. Because of the uncertainties in the measured parameters
the allowed solution is not confined to the line but it can lie
anywhere within the shaded box.
\label{fig:illust}}
\end{figure}

The single number {$\cf$} gives the fraction of all photons at a given
wavelength that pass through absorbers along the line of sight.
However, at the position of these associated absorption lines in the
spectrum, the absorbed photons have two significant sources, the
continuum source and the BELR.  The continuum source is likely to be
significantly smaller than the BELR, but the geometry and relative
position of the BELR is unknown. In this section, we continue to
assume that the photons from these two regions pass through the same
absorbers, i.e. that the optical depth $\tau$ is the same along the
paths to the observer from the continuum source and from the BELR.
However, we now consider the different coverage fractions, {$\cc$} and
{$\celr$}, that can apply for the continuum source and the BELR.
First we define, $W = F_{\rm elr}/F_{\rm c}$, as the ratio of the
broad emission--line flux to the continuum flux at the wavelength of
the narrow absorption line.  Then we can write the normalized flux as:
\begin{equation}
R = 1 - {{(\cc + W \celr) (1-e^{-\tau})} \over {(1+W)}} \quad .
\end{equation}
By the same optical depth scaling argument as in the previous section, this
reduces down to:
\begin{equation}
{{\cc + W \celr} \over {1 + W}} = {[(R_{\rm r}(v)-1)]^2 \over
{R_{\rm b}(v)-2R_{\rm r}(v)+1}} = \cf \quad ,
\end{equation}
where we have assumed that underlying, unabsorbed continuum plus line
flux is the same in both transitions, i.e., that the value of $W$ is
the same for both members of the doublet. In the Appendix we calculate
that {$\cc$} or {$\celr$} will change by at most 15\% if a doublet is
on the steep slope of an emission line due to a differing $W$.
Equation~(5) shows that the effective coverage fraction, {$\cf$}, can
be considered as an average coverage fraction weighted according to
the flux from each source. Of course, with the available information,
the continuum source and BELR coverage fractions cannot be determined
independently of each other. It is possible, nevertheless, to place
interesting constraints on them.  Given the value of $\cf$ and $W$,
and their $1\sigma$ uncertainties, equation~(5) defines a region in
the $\cc$--$\celr$ parameter plane where the solution must
lie. Examples of such regions are illustrated in
Figure~\ref{fig:illust}. It is interesting to note that, if the upper
boundary of this allowed region intersects one of the axes at a value
below 1, then the value of the corresponding coverage fraction must be
less than unity (at the $1\sigma$ confidence level). Such a situation
can occur if the narrow absorption doublet happens to fall on the
high--velocity wing of an emission line where the underlying flux is
dominated by the continuum (i.e., when $W$ is small). Because of this
technical requirement, we cannot draw any conclusions about the
preferred velocity distribution of systems that cover the continuum
source only partly.

Using the observed values of $W$ from published low--resolution spectra
(\cite{sbs88}; \cite{ss91}) and the measured values of $\cf$ we have
derived constraints on the coverage fractions of the two distinct
sources. These limits, along with the ingredients needed to compute
them, are listed in Table~\ref{tab:covfac2} for each system and
transition showing the signature of partial coverage. In particular,
in the last two columns of Table~\ref{tab:covfac2} we list the range
of allowed values of {$\cc$} and {$\celr$} as deduced from the range
of allowed solutions of equation~(5).  We emphasize that the limits on
{$\cc$} and {$\celr$} do not hold independently of each other since
the solution should also satisfy equation~(5). In practice these
limits correspond to the extreme corners of the solution boxes as
illustrated in Figure~\ref{fig:illust}.  The constraints on {$\cc$}
and {$\celr$} are also represented graphically by the vertical extent
of shaded boxes and vertical bars, respectively, in
Figures~\ref{fig:q0450f2}--\ref{fig:q1329f2}.

In most of the absorption components listed in
Table~\ref{tab:covfac2}, either the continuum source or the BELR could
be fully covered.  Notable exceptions are the {\SiIV} line of
Q~$0450-132$ and the highest--velocity component of its {\NV} line (at
$\Delta v=-1880$~{\kms}; see Figure~\ref{fig:q0450f2}), as well as
the red {\CIV} component in PG~$1222+228$ (at $\Delta v=+1587$~{\kms};
see Figure~\ref{fig:q1222f2}).  In these cases {$\celr$} is not
constrained, but the continuum source cannot be fully covered under
any conditions. We derive $0.59 < {\cc} < 0.88$ and ${\cc} < 0.84$,
respectively, in the case the former object and $0.12 < {\cc} < 0.52$
in the case of the latter object. In the Appendix we consider the
possibility that the photons from the continuum source and the BELR
pass through different absorbers, i.e. $\tau_c \ne \tau_{elr}$.  This
could be possible depending on the sizes of absorbing structures and
their spatial positions relative to the emitting regions, but we show
in the case of the redward component of PG~$1222+228$ partial coverage
of the continuum source is required regardless of the assumed $\tau_c$
and $\tau_{elr}$.  The $\Delta v=+1587$~{\kms} component of the
PG~$1222+228$ system also requires a partially covered continuum
source, with a larger continuum coverage fraction. It is interesting
that the range of constraints $0.51 < \cc < 0.92$ does not overlap
with that of the $\Delta v=+1469$~{\kms} component.

\subsection{Application of the Refined Method to Additional Quasars With
Published Data}

There are three other quasars with intrinsic {\CIV} narrow line
absorbers observed with HIRES/Keck~I and presented in the literature
in sufficient detail to derive {$\cf$}, {$\cc$}, and {$\celr$}.  These
are PKS~$0123+237$ (radio--loud; \cite{bs97}), Q~$0150-203$ (UM 675;
radio--quiet; \cite{ham97a}), and Q~$2343+125$ (radio--quiet;
\cite{ham97b}). The results we obtain by applying the method of the
previous section to these objects are listed in Table~\ref{tab:covfac}.

The NAL systems of PKS~$0123+237$ were discussed by Barlow and Sargent
(1997\nocite{bs97}).  The velocity structure of
the NAL profile of PKS~$0123+237$ is similar to that of Q~$0450-132$
but in the former object the effective coverage fraction is different
for different transitions. In the case of Q~$2343+125$ (\cite{ham97b})
the NALs are found at a very large velocity relative to the
emission--line peak ($\Delta v=-24,000$~{\kms}) but still there is
considerable evidence that they are intrinsic: in addition to showing
the signature of partial coverage they also happen to be variable.
The continuum coverage fraction can be constrained to be very small
($<0.2$) because the emission--line contribution at the velocity of
absorption is small ($W = 0.1$; see Table~\ref{tab:covfac}).

Together with the NAL systems presented in the previous section, the
systems analyzed here comprise the database of available
high--resolution spectra of NALs.  All six intrinsic NAL systems have
{\CIV} absorption profiles with velocity widths of $100$--$400$~{\kms}
and have obvious sub--structure (see, for example,
Figure~\ref{fig:q0450f1}).  The NAL profiles suggest that there are
discrete absorbing structures (though not necessarily discrete
``clouds'') at different positions in velocity space.

\section{Discussion}

\subsection{Implications of the Observational Results}

Since the six quasars of Table~\ref{tab:qso} were drawn from a sample
that was selected based on the properties of {\it intervening} {\MgII}
absorbers, they can be regarded as a collection which is unbiased
towards the detection of {\it associated} {\CIV} absorption lines. As
such we can use it to estimate roughly how frequently associated
{\CIV} NALs are found in radio--loud and radio--quiet quasars.
Associated absorption lines were detected in all three radio--quiet
objects, and, in two cases, the lines were shown to be intrinsic based
on the signature of partial coverage. The two NAL systems in
Q~$1213-003$, which did not show the signature of partial coverage,
could plausibly be intervening. One would expect to find 2 intervening
systems within 5000~{\kms} in a sample of six quasars, based on the
known density of {\CIV} systems at $z=2$ and the redshift path that we
have observed.  On the other hand, only one out of the three
radio--loud objects showed an associated NAL system which also turned
out to be intrinsic.

Our results are consistent with previous studies 
(\cite{foltz86}; \cite{and87}) that find that associated NALs
are fairly common in both radio--loud and radio--quiet quasars.
However, as Foltz \etal~(1986\nocite{foltz86}) pointed out, there
could be a systematic difference between the strengths of associated
NALs found in radio--loud and radio--quiet quasars.  Strong
NALs ($>1.5$~{\AA}) prefer radio--loud hosts, while weak NALs show
no such preference.  None of the five associated systems detected
in our survey are strong, thus we cannot re--address this issue.
It should be noted that although the associated systems
that we report here are fairly weak, in three our of four quasars they
are still within the detection limits of the 
Foltz \etal~(1986\nocite{foltz86}) and Anderson \etal~(1987) surveys.
However, our more sensitive survey shows that in some cases
(Q~$0002+051$ and Q~$1421+331$) NALs are {\it not} detected down to
a $5\sigma$ limit of rest--frame equivalent width, $REW < 0.05$~{\AA}.  
This implies that there
may be some lines of sight that do not pass through NAL gas.

More than 10 intrinsic NAL systems reported in
the literature appear in radio--quiet quasars (\cite{ham97c};
\cite{bhs97}; Tripp, Lu, \& Savage 1997\nocite{tripp}). It is
interesting that, although Foltz \etal~(1986\nocite{foltz86}) and
Anderson \etal~(1987\nocite{and87}) find associated NALs in about 70\%
of radio--loud quasars, only very few of these systems have been proven
rigorously to be intrinsic: two systems from the sample of Foltz
\etal~(1986) show signs of variability (3C~$205$ and the ``mini--BAL''
in PHL~$1157+0128$; \cite{ald97}). Moreover, some fraction of
associated NAL systems are bound to be intervening systems by chance;
in fact 2 out of the 12 associated NAL systems in the sample of Anderson
\etal~(1987\nocite{and87}) are expected {\it a priori} to be
intervening systems.
A description of the properties of {\it intrinsic} NALs in
radio--loud and radio--quiet quasars
has not yet been established.  A systematic survey at very high
spectral resolution is needed to clarify the situation.

Another important observational result is that, in at least two
quasars from our collection, the NAL gas can be demonstrated to cover
the {\it continuum source} only partially. One of these quasars,
Q~$0450-132$, is radio--quiet, while the other, PG~$1222+228$, is
radio--loud. Partial coverage of the continuum source may be more
common than these two cases suggest but we have no way of determining
this yet.  The implication of this result is not absolutely clear
because the reason for the apparent partial coverage is not known. One
possibility is that the absorbing medium is clumpy and the clump sizes
are comparable to or smaller than the size of the continuum
source. Another possibility, however, is that the continuum source is
completely covered by the absorber but a scattering medium makes it
possible for continuum photons to bypass the absorber and get to the
observer without suffering any absorption. We expand on this issue in
the next section where we consider possible explanations for the
observed partial coverage.

\subsection{Interpretation of Effective Partial Coverage}

Central to interpreting these results is knowledge of the mechanism
that can give rise to effective coverage fractions less than unity.
Following our preliminary discussion of this subject in \S3, three
mechanisms can fill in saturated, non--black troughs, or more generally
produce discrepant optical depth ratios for doublet or multiplet
transitions: 1) a source function, 2) scattering of background photons
back into the light path, 3) true (i.e., geometric) partial coverage
of the continuum and/or BEL source by the absorbing medium.  The
present data do not permit us to distinguish among these
possibilities, but we discuss possible tests.  These tests involve
polarization measurements, consideration of the coverage fraction in
different transitions, and correlating time variability with coverage
fractions.

One possible source function explanation involves a non--occulted
emission source which contributes additional photons to the light path
after absorption of the continuum and BEL photons has occurred,
essentially ``diluting'' the absorbed spectrum. Wampler
\etal~(1995\nocite{wam0059})
determined that an unabsorbed thermal continuum characterized by
$T=17,000$~K could explain the residual intensity in the saturated
troughs of various transitions in the BAL system of Q~$0059-2735$. In
fact, other intrinsic NAL systems also show variations in $\cf$ from
transition to transition, namely UM675 (\cite{ham97a}) and
PKS~$0123+257$ (\cite{bs97}). The mechanism for production of the
additional source function photons is unknown, but its contribution
should vary smoothly with wavelength to produce the observed
effect. Another manifestation of a source function is line emission by
the same gas that is responsible for the absorption. This version of
the scenario is not viable, however, since it would not have the
desired effect: if there were an emission component filling in the
unresolved absorption troughs it would result in an effective coverage
fraction {\it greater that unity}.

Scattering of photons back into the light path is certainly a
plausible explanation.  An indirect argument for a scattering
explanation is that it is known to be important in filling in the
troughs of {\it broad} absorption lines. Spectropolarimetric
observations of many BAL quasars (\cite{ogle97}; \cite{broth97};
\cite{cohen95}; Glenn, Schmidt, \& Foltz 1994\nocite{glenn94}; 
\cite{good95}; \cite{hines95}; \cite{schmidt97}) show that the 
polarization level in saturated non--black absorption troughs is higher
than in the continuum.  This by no means proves that scattered light
makes a large contribution for all NAL absorbers, since the lines of
sight through these absorbers could well be different than in
BALs. Spectropolarimetric observations of partially covered NAL
systems are needed to investigate the role of scattering.  Until such
observations are carried out, we can consider clues provided by the
relationship between time variability and partial coverage. The
implications of the observational clues, however, depend on the
picture that one adopts for the scatterers. If the scatterers comprise
an extended collection of clumps or a quasi--uniform medium enveloping
the BELR, then the scattering contribution is unlikely to vary
substantially over time scales of the order of the dynamical time of
the BELR gas. The strength of absorption, on the other hand, can vary
as a result of changes in ionization state of the absorber or motion
of the absorbing gas across the line of sight. One alternative picture
is that of a discrete clump acting as a mirror to scatter continuum
and/or emission--line photons in the direction of the observer without
their passing through the absorber.  In this picture the amount of
scattered light can vary on the same time scale as the absorbing
column density making a unique interpretation of the variability time
scale virtually impossible. Another alternative picture is that the
scatterers are close to our line of sight (they could be mixed with
the absorbers or they could form a tenuous, ionized atmosphere round
individual clumps in the absorber). In this case the light is
scattered in the forward direction and its fractional polarization
does not change. In other words, even though scattering is the culprit
in this scenario, it does not betray itself by its polarization
signature.

At present there is one NAL system where variability provides strong
evidence {\it against} a scattering mechanism. The $\Delta
v=-24,000$~{\kms} {\CIV} NAL system toward Q~$2343+125$ has an
effective coverage fraction that varies substantially from one epoch
of observation to another, and the effective coverage fraction is
smaller when the absorption lines are weaker (\cite{ham97a}). This
argues against a scattering picture of the above type, in which
scattered light would fill in a larger fraction of the trough of a
weaker line.  From the time variability and the partial coverage of
the continuum source for this system, Hamann {\etal} (1997b) derived
the requirement that the absorbing clouds are quite small ($\sim
0.01$~{\rm pc}) and are located close to the continuum source. If we
adopt this interpretation, then true partial coverage of the continuum
source is the most plausible explanation. This suggests that time
variability and true partial coverage occur together because both are
related to small absorbers close to the continuum and/or BEL source.
The variations can result from either motion of the clouds or from
changes in the sizes of the multiphase layers of the absorber.
Discriminating between these two possibilities is challenging and
requires a geometric model of the absorbers and the sources of
ionization, which can be constrained by observations of multiple
transitions.

\subsection{A General Picture of Intrinsic NALs}

Narrow absorption lines are ubiquitous, appearing in all types of
broad--line AGNs, namely in Seyfert~1 galaxies and broad--line
radio galaxies as well as in radio--loud
and radio--quiet quasars. If we take the frequency of narrow {\it
associated} absorption lines as a rough indication of the coverage
fraction of the absorbers (as seen by the continuum and/or
emission--line source) then this fraction appears to be very high. It
is about 50\% in Seyfert~1 galaxies (Crenshaw 1997), perhaps about
70\% in radio--loud quasars (Foltz \etal~1986; Anderson \etal~1987),
and apparently also very high in radio--quiet quasars based on our small
collection.

The observational clues available at the moment, although tantalizing,
do not point clearly to a specific interpretation. Because NALs are
seen in many different types of active galaxies and seem to be quite
common, their relation to BALs is unclear.  Unlike BALs, NALs are
common in radio--loud quasars and in Seyfert 1's. It is plausible that
NALs with large velocities and those that are quite broad and without
structure (``mini--BALs'') are related to the BAL phenomenon
(\cite{ham97c}). This hypothesis is supported by the fact that large
outflow velocities ($>$10,000~{\kms}), such as in Q~$2343+125$,
Q~$0935+417$ (\cite{ham97c}), and Q~$1700+64$ (\cite{tripp}), have
only been observed in radio--quiet quasars, as have the majority of
BALs. Radio--loud quasars on the other hand have mostly narrow,
``associated'' absorption lines.  Finally, associated NALs found in
Seyfert~1 galaxies show limited observational evidence for an
intrinsic nature. Some have covering factors close to unity
(\cite{cren97}), and in two cases, NGC~3783 and NGC~3516, the NALs
vary on a short time scale (\cite{korat96}; \cite{shields97}).
Interestingly enough NALs in Seyfert~1 galaxies are always blueshifted
relative to the BELs, unlike NALs in quasars which are sometimes
redshifted relative to the BELs. Based on the above summary, a
reasonable model or scenario for the NAL gas should explain why NALs
are so common (in other words, why the absorbing gas covers such a
large solid angle relative to the continuum source) and why they are
sometimes observed to be redshifted and sometimes blueshifted relative
to the peaks of the broad emission lines (i.e. apparent infall 
as well as outflow).

\begin{figure}[th]
\figurenum{11}
\plotfiddle{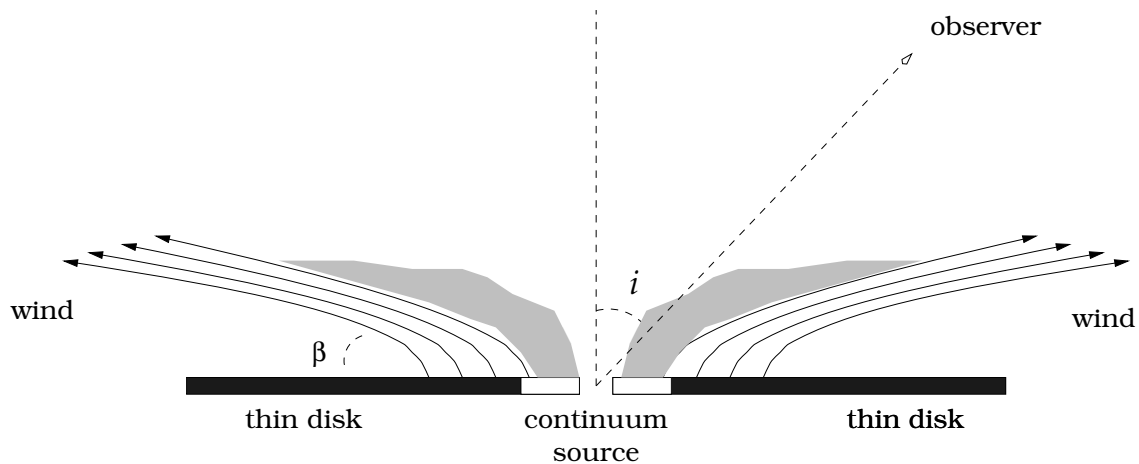}{3.0in}{270}{70}{70}{-272}{350}
\protect\caption{A schematic illustration of the geometry of the fast (BAL)
accretion--disk wind of Murray \etal~(1995) and a plausible location
of the NAL gas (grey--shaded region). The arrows show the streamlines
of the fast wind. The NAL gas could form an atmosphere or an interface
between the fast wind and the surrounding medium (indicated in grey).
The observer's line of sight is inclined at an angle $i$
relative to the axis of the disk, while the opening angle of the fast
wind relative to the disk plane is $\beta$. If the NAL gas is
outflowing along with the fast wind and is seen in absorption against
it, then the NALs from the far side of the outflow will appear
redshifted to the observer if $\beta < i$.
\label{fig:cartoon}}
\end{figure}

It is obviously impossible to construct a unique unifying model to
explain these observational facts.  Nevertheless, we venture to
speculate on how NALs can fit into the accretion--disk wind picture of
Murray {\etal} (1995\nocite{mur95}), which was originally meant to
explain BALs. Their possible relatives, the high velocity NALs, could
originate in a different phase of the same region or in an atmosphere
just outside the wind.  Sometimes both NALs and BALs arise in the same
quasar, at different velocities, so it seems plausible that NAL phases
could exist nearby the BAL region.  The broader, smoother, mini--BALs
may also be related to this type of wind flow. In this context, it is
extremely interesting that in some cases the NALs are redshifted
relative to the {\it broad} emission--line peak. Of course this
relative redshift does not necessarily imply infall towards the
central engine. The systemic redshift is best determined from the {\it
narrow} emission lines, which, however, have not been observed in
these particular objects. Because the peaks of the broad emission
lines can be blueshifted relative to the systemic redshift, it is
possible that the observed NALs are also blueshifted relative to the
systemic redshift.  At any rate, all four of the NAL systems in
radio--loud quasars that have been demonstrated to be intrinsic, i.e.,
PG~$1222+228$ (this paper), Q~$0123+237$ (\cite{bs97}), $3$C~$205$
(\cite{ald97}), and PHL~$1157+014$ (\cite{ald97}), have NALs that {\it
appear to be} redshifted with respect to the peaks of the broad
emission lines.  In Figure~\ref{fig:cartoon}, we show schematically a
side view of the geometry of the accretion--disk wind of Murray {\etal}
(1995) and a plausible location for the NAL gas.  For specific
orientations of the observer (such that $\beta < i$, where $i$ is the
inclination of the disk relative to the observer and $\beta$ is the 
opening angle of the wind,
as shown in the figure), the NALs originating in the far side of the
disk can appear redshifted, assuming that the NAL gas is outflowing.
Since the NALs are sometimes deeper than the net emission--line level
on which they are superposed, the absorbing gas must intercept
continuum photons as well. To accommodate such an effect in this
picture we must postulate that the outflowing NAL gas originates very
close to the center of the disk so that it can cover the continuum
source, i.e., it fills the region between the fast disk wind
and the disk axis. We emphasize, however, that alternative
pictures are also plausible. For example, the NAL gas could be located
on the near side of the disk, in the direction of the
observer. Because the wind is rotating as well as outflowing, the net
velocity vector of the outflowing gas can point away from the observer
depending on the azimuth, in which case the resulting absorption lines
will also appear redshifted.

In the context of the speculative geometric picture outlined above,
the difference between AGN subclasses may be related to the properties
of the fast wind. For example, if the opening angle of the wind is
very small, the wind effectively ``hugs'' the accretion disk and the
likelihood of our line of sight going through it is very small. Hence
BALs would not be observed in such objects but NALs could still
appear. We speculate further that the difference between
radio--quiet quasars on one hand and radio--loud quasars and Seyferts on
the other is the opening angle of the wind, which may in turn be a
consequence of the luminosity of the object (relative to the
Eddington luminosity), of the shape of the ionizing continuum, or
of the combination of these two properties.
This could also affect the strength of NALs in the two cases as
observed by Foltz \etal~(1986) for radio--loud vs. radio quiet quasars.
Finally, we note that if any hot electrons happen to be located in the
direction of the axis of the disk, they can act as scatterers and
redirect continuum photons towards the observer. Through this effect,
the absorption lines can be ``diluted'' creating the impression that
the absorber only partly covers the continuum source.


\section{Summary and Conclusions}

In this paper we reported the discovery of five narrow ``associated''
($\vert\Delta v\vert<5000$~{\kms}) absorption--line systems in the
spectra of four quasars. These quasars were observed in the course of
a survey for intervening {\MgII} absorbers, in which the {\CIV}
emission lines of six objects happened to fall in the observed
spectral range.  As such, this ``mini--sample'' of six quasars is
unbiased towards the presence of narrow, intrinsic {\CIV} absorption
lines.  Three of the absorption systems were demonstrated to be
intrinsic, because the relative strengths of the members of the
{\CIV} doublet required effective partial coverage of the continuum
and/or emission--line sources. The two systems toward Q~$1213-003$
could not be demonstrated as intrinsic because their coverage
fractions were consistent with unity.

We carried out a detailed analysis of the coverage fractions of the
three intrinsic systems that we have found: Q~$0450-132$,
PG~$1222+228$, and PG~$1329+412$. The analysis consisted of
determining the effective coverage fraction of the background
source(s) by the absorber according to the method of Barlow \& Sargent
(1997\nocite{bs97}) as well as a refinement of this method to treat
the coverage fractions of the background continuum and emission--line
sources separately. The latter technique was also applied to three
additional intrinsic NAL systems for which sufficient information was
available in the literature: PKS~$0123+237$, UM~$675$, and
Q~$2343+125$.  This provided a total of six intrinsic NAL systems
studied at high resolution, and among the six there was a very wide
range of properties.  The derived {\CIV} effective coverage fractions
ranged from $\cf \sim 0.1$ to nearly unity, and in some systems varied
across the absorption profiles.  The velocities of the absorbing
clouds, relative to the quasar, ranged from outflow at $24,000$~{\kms}
to apparent infall at $1500$~{\kms}.  In three systems, Q~$0450-132$
and PG~$1222+228$ (this study) and Q~$2343+125$ (\cite{ham97b}), the
data require that the covering factor of the {\it continuum} source is
less than unity ($\cc < 1$, according to our formulation presented in
\S4.2).  The two systems are quite different from each other:
Q~$2343+125$ is radio--quiet and its absorption system is at an outflow
velocity of $24,000$~{\kms}, while PG~$1222+228$ is radio--loud and its
absorption system is redshifted relative to the BEL region by
$1500$~{\kms}.

We have discussed mechanisms which can give rise to a covering factor
that is less than unity but we have not been able to select a favorite
based on the information that is currently available from
observations. We have also speculated that the NAL gas may be related
to the fast outflows that manifest themselves as BALs.  If this is
indeed the case, then the fact that NALs are quite common in radio--loud
quasars and in Seyfert~1 galaxies may imply that these types of AGN
also harbor fast accretion--disk winds, whose absorption signature
(BAL) is unobservable. Further observations, and more specifically
systematic surveys of the various classes of active galaxies and
quasars, are needed to test these ideas.  Some of the key goals of
these surveys should be to: 1) establish just how common {\it
intrinsic} NALs are in radio--loud/quiet quasars and in Seyfert 1's; and
2) check if non--unity covering factor and time variability
occur together as a clue toward understanding the spatial distribution
of the NAL clouds. In addition spectropolarimetric observations are
sorely needed to establish whether scattering or true partial coverage
by small absorbers is responsible for effective covering factors less
than unity.

\acknowledgments

We thank Jules Halpern as well as the anonymous referee for useful
suggestions.  This work was supported by the National Science
Foundation Grants AST--9529242 and AST--9617185, and by NASA Grant
NAG5--6399.  During the early stages of this work M.E. was based at the
University of California, Berkeley and was supported by a Hubble
Fellowship (grant HF--01068.01--94A from Space Telescope Science
Institute, which is operated for NASA by the Association of
Universities for Research in Astronomy, Inc., under contract
NAS~5--26255). We are grateful to Steven S. Vogt for the great job he
did building HIRES.

\section*{APPENDIX}

\appendix

\section{More General Forms of the Partial Coverage Calculation}
\label{app:general}

Here we consider the consequences of two simplifying assumptions that
we made in the calculation of the coverage fraction in \S4.2: (1) the
emission line flux is the same for both members of the doublet; and
(2) the optical depths toward the continuum source and the BLR are the
same.

\subsection{Underlying Emission--Line Flux Differs from Stronger to Weaker
Transition}

We have assumed that the ratio of the emission line to the
continuum flux is the same in the blue and red components of
a doublet, i.e. that $W_{\rm b} = W_{\rm r} = W$. This assumption may
lead to inaccuracies in the derivation of the coverage fraction when
the absorption falls on the steep part of an emission feature so that
$W_{\rm b}$ and $W_{\rm r}$ may differ by $10$--$15$\%.  In this more
general case we have
\begin{equation}
{{(1 - R_{\rm b}) (1+W_{\rm b})} \over {\cc + W_{\rm b} \celr}} =
{{2(1-R_{\rm r})(1+W_{\rm r})} \over {\cc + W_{\rm r} \celr}} -
\left [ {{(1-R_{\rm r})(1+W_{\rm r})} \over {\cc+W_{\rm r} \celr}} \right ]^2
\quad ,
\end{equation}
which can be reduced to a quadratic equation in $\celr$ if $\cc$ is specified
(or in $\cc$ if $\celr$ is specified).

To see how the assumption used previously will affect our
results, consider the quasar Q~$0450-132$ which has its absorption
on the steepest area of its emission feature.  In this
case, using $W_{\rm r}=0.83$ and $W_{\rm b}=0.7$ changes {$\cc$} or
{$\celr$} only by {$12-15$}\%.

\subsection{Different Absorbers Cover Continuum and BEL Sources}

Here we examine the more general case of incomplete coverage of the
two distinct sources of photons (continuum source and BEL) where
the light from the two sources is allowed to pass through different
absorbing clouds (or significantly different regions of the same
cloud). 
Let {$\tauc$} and {$\taue$} be the optical depths along the lines of sight 
from observer to
continuum source and from observer to BEL source. If the clouds occult a
fraction {$\cc$} and a fraction {$\celr$} of the lines of sight toward each
source, then the normalized residual intensity is:
\begin{equation}
R = {{(1-\cc) + \cc e^{-\tauc} + (1-\celr) W + \celr W e^{-\taue}} \over
{(1+W)}} \quad ,
\end{equation}
where $W$ is the emission line flux in continuum units. In conjunction
with optical depth scaling laws, this gives a system (for an
$n$--transition multiplet) of $3n-2$ equations ($n$ residual intensity
equations, $n-1$ continuum optical depth relations, $n-1$ ELR optical
depth relations) and $2(n+1)$ unknowns ($2$ coverage fractions and
$2n$ optical depths). This can, in principle, be solved for a
multiplet with at least four transitions.

\begin{figure}[th]
\figurenum{4}
\plotone{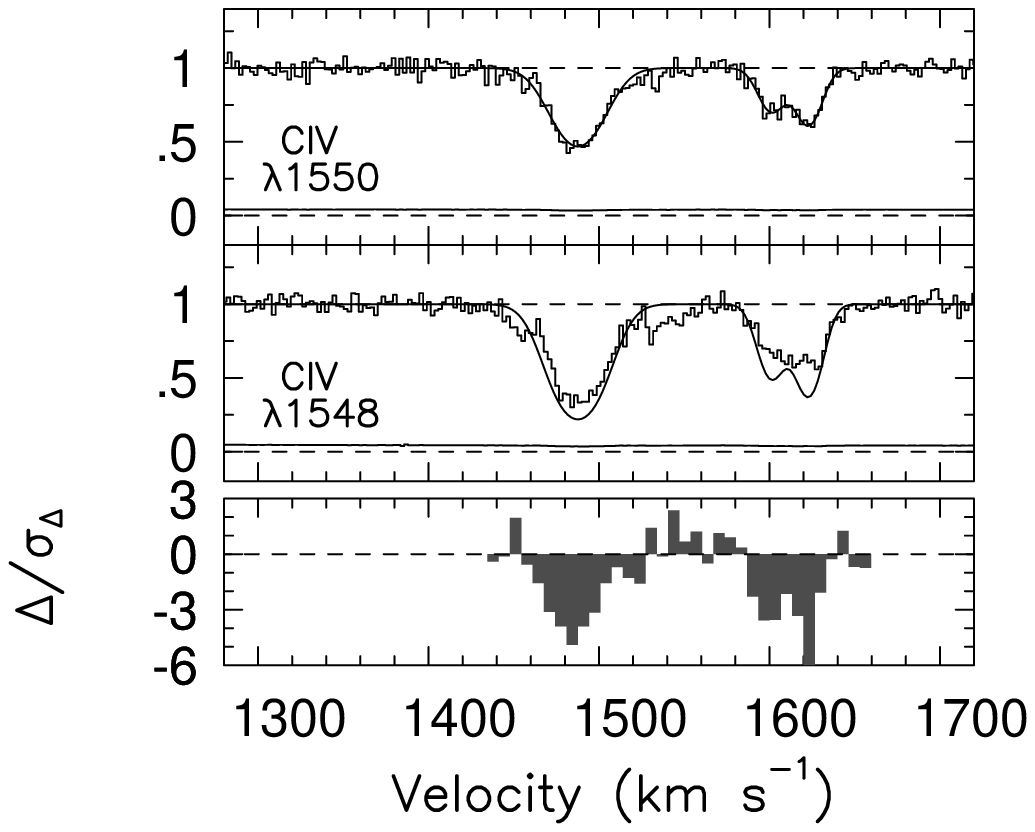}
\vglue -0.9in
\protect\caption{Same as Figure~1, but for the {$\Delta v=+1482$~\kms} system toward
{PG~$1222+228$}. The obvious discrepancy between the atomic physics
prediction and the data in this particular case is indicative of
partial coverage.
\label{fig:q1222f1}}
\end{figure}

With this more general case in mind we now reconsider the issue of
partial coverage of the continuum source in PG~$1222+228$ to see if
this interesting result still applies. At the position of the $\Delta
v=+1587$~{\kms} component of the {\CIV} absorption
(Figure~\ref{fig:q1222f1}) the continuum contributes $\sim 77$\% of
the total flux (i.e., $W = 0.3$), and the BELR only $23$\%. The
observed flux at that position is 0.6 in normalized units.  Therefore
it is not possible to have zero coverage of the continuum source
because then the observed flux would be {\it at least} 0.77 no matter
what the values of {$\celr$} and {$\taue$}. If ${\cc} = 1$ then $0.24
< {\tauc}(\lambda1550) < 0.73$.  The lower limit is provided by the
requirement that the continuum photons alone (those that pass through
the absorber) do not produce too much flux.  The upper limit is the
minimum contribution from unimpeded continuum photons since the BEL
can contribute at most a flux of 0.3. Over that range of {$\tauc$},
the continuum photons that pass through the absorber produce a flux
ranging from 0.6 to 0.3 at the position of {\CIV}~$\lambda$1550 and
from 0.48 to 0.18 at the position of {\CIV}~$\lambda$1548.  In either
case, the {\it observed} flux at 1548~{\AA} is just slightly smaller
than at 1550~{\AA} (0.60 as compared to 0.62).  To be consistent with
these data, the contribution to the flux from the BEL photons would
have to be larger at 1548~{\AA} than at 1550~{\AA}. The contribution
from photons that do not pass through absorbers is the same for the
two transitions, and the contribution from photons at 1550~{\AA} that 
make it through the absorber should be equal or larger than at
1548~{\AA}.  This is the opposite of what is required.  Thus the
continuum source cannot be fully covered at any {$\tauc$}.  Since
${\cc} \ne 0$ and ${\cc} \ne 1$ we conclude that the continuum source
must be partially covered regardless of assumptions about {$\tauc$}
and {$\taue$}.



\begin{deluxetable}{rlccrrcl}
\tablewidth{0pc}
\tablenum{1}
\tablecaption{Summary of Quasars Properties}
\tablehead
{
\colhead{Object} &
\colhead{Other Names} &
\colhead{$z_{\rm em}$ \tablenotemark{a}} &  
\colhead{$m_{\rm V}$} & 
\colhead{$P_{\rm\, 5~GHz}$\tablenotemark{b}} & 
\colhead{$R$ \tablenotemark{c}} &
\colhead{Radio Class} &
\colhead{Ref.\tablenotemark{d}}\nl
\colhead{} &
\colhead{} &
\colhead{} &
\colhead{(mag)} & 
\colhead{(W~Hz$^{-1}$)} &  
\colhead{} & 
\colhead{} & 
\colhead{}
} 
\startdata
Q~$0002+051$  & UM018, PHL 0650 & 1.899 & 16.2 & $ 2\times 10^{27}$ & 110\phantom{.0} & loud  & 1,2 \nl
Q~$0450-132$  &                 & 2.253 & 17.5 & $<4\times 10^{25}$ & $<5$\phantom{.0}& quiet & 3 \nl
Q~$1213-003$  & UM485           & 2.691 & 17.0 & $<5\times 10^{25}$ & $<3.5$          & quiet & 3 \nl
PG~$1222+228$ & Ton 1530        & 2.040 & 15.5 & $ 1\times 10^{26}$ & 3.8             & loud  & 4,5 \nl
PG~$1329+412$ &                 & 1.937 & 16.3 & $ 5\times 10^{24}$ & 0.3             & quiet & 4 \nl
Q~$1421+331$  & Mrk 679         & 1.906 & 16.7 & $ 2\times 10^{26}$ &  17\phantom{.0} & loud & 6 \nl
\enddata
\tablenotetext{a}{Emission redshift as determined from the peak of the broad
{\CIV} emission line (\cite{sbs88}; \cite{ss91})}
\tablenotetext{b}{Radio power at a rest frequency 5~GHz (see text in \S2 for more details).}
\tablenotetext{c}{Ratio of rest-frame 5~GHz-to-4400~\AA\ flux densities.}
\tablenotetext{d}{{\sc References to Radio Observations.--}
(1) 87GB survey; \cite{87gb};
(2) \cite{gps};
(3) NVSS survey; \cite{con98};
(4) \cite{kel94};
(5) \cite{barv96};
(6) FIRST survey; \cite{first}}
\label{tab:qso}
\end{deluxetable}

\clearpage

\begin{deluxetable}{rcccrcc}
\tablewidth{0pc}
\tablenum{2}
\tablecaption{Observation Log and Absorption Line Propertires}
\tablehead
{
\colhead{} &
\multicolumn{3}{c}{Observation Log} &
\multicolumn{3}{c}{Absorption-Line Properties} \nl
\colhead{} &
\multicolumn{3}{c}{\hrulefill} &
\multicolumn{3}{c}{\hrulefill} \nl
\colhead{Object} &
\colhead{UT Date \tablenotemark{a}} & 
\colhead{Exposure} & 
\colhead{$\lambda$ Range} &
\colhead{$\Delta v$\tablenotemark{b}} &
\colhead{$REW$\tablenotemark{c}} &
\colhead{$EW_{\rm min}$\tablenotemark{d}} \nl
\colhead{} &  
\colhead{} & 
\colhead{(s)} & 
\colhead{(\AA)} &
\colhead{(\kms)} &
\colhead{(\AA)} &
\colhead{(m\AA)} 
} 
\startdata
Q~$0002+051$  & 1994 Jul  5 & 2700 & 3656--6079 & \nodata & \nodata         & 50 \nl
Q~$0450-132$  & 1995 Jan 24 & 5400 & 3987--6425 & $-2011$ & $1.028\pm0.006$ & 30 \nl
Q~$1213-003$  & 1995 Jan 24 & 5200 & 5008--7357 & $-1566$ & $0.108\pm0.003$ & 11 \nl
              &             &      &            & $  -60$ & $0.149\pm0.004$ & 13 \nl
PG~$1222+228$ & 1995 Jan 23 & 3600 & 3811--6305 & $+1482$ & $0.268\pm0.005$ & 17 \nl
PG~$1329+412$ & 1996 Jul 18 & 6300 & 3766--5791 & $ +314$ & $0.371\pm0.009$ & 51 \nl
Q~$1421+331$  & 1995 Jan 23 & 3600 & 3819--6317 & \nodata & \nodata         & 50 \nl
\enddata
\tablenotetext{a}{The 1996 July exposures were severly affected by cloud cover.}
\tablenotetext{b}{Velocity offset of the centroid of the {\CIV} associated absorption 
profile (\cite{sbs88}; \cite{ss91}). A negative value denotes a blueshift.}
\tablenotetext{c}{The rest--frame equivalent width of the C\kern 0.1em{\sc iv}~$\lambda 1548$ absorption line.}
\tablenotetext{d}{The minimum measurable equivalent width ($5\sigma$ limit; \cite{ltw87}) 
in the velocity range of interest.}
\label{tab:obs}
\end{deluxetable}

%
%

\clearpage

\begin{deluxetable}{lccccc}
\tablewidth{6in}
\tablecolumns{6}
\tablenum{3}
\tablecaption{Coverage Fractions Averaged over Small Velocity Bins\tablenotemark{a}}
\tablehead
{
\colhead{Ion} &
\colhead{$\langle v\rangle$\tablenotemark{b}} &
\colhead{$\cf$} &
\colhead{W} &
\colhead{$\cc$\tablenotemark{c}} &
\colhead{$\celr$\tablenotemark{c}} \nl
\colhead{} &
\colhead{(\kms)} &
\colhead{} &
\colhead{} &
\colhead{} &
\colhead{} 
}
\startdata
\cutinhead{Q~$0450-132$} 
\CIV  & $-2109$ & $0.83\pm0.02$ & $0.75\pm0.02$ & $>0.65$       & $>0.53$       \nl
      & $-2057$ & $0.61\pm0.08$ &               & $>0.35$       & $>0.10$       \nl
      & $-2014$ & $0.93\pm0.02$ &               & $>0.83$       & $>0.72$       \nl
      & $-1940$ & $0.75\pm0.06$ &               & $>0.45$       & $>0.22$       \nl
      & $-1870$ & $0.68\pm0.10$ &               & $>0.24$       & unconstrained \nl
\SiIV & $-2108$ & $0.72\pm0.08$ & $0.10\pm0.02$ & 0.59--0.88    & unconstrained \nl
\NV   & $-2051$ & $0.48\pm0.24$ & $0.50\pm0.02$ & unconstrained & unconstrained \nl
      & $-2013$ & $0.85\pm0.06$ &               & $>0.68$       & $>0.16$       \nl
      & $-1948$ & $0.69\pm0.29$ &               & $>0.09$       & unconstrained \nl
      & $-1880$ & $0.36\pm0.20$ &               & $<0.84$       & unconstrained \nl
\cutinhead{PG~$1222+228$}
\CIV  & $+1469$ & $0.67\pm0.04$ & $0.30\pm0.02$ & 0.51--0.92    & unconstrained \nl
      & $+1587$ & $0.36\pm0.04$ &               & 0.12--0.52    & unconstrained \nl
\cutinhead{PG~$1329+412$}
\CIV  &  $+273$ & $0.81\pm0.04$ & $0.70\pm0.02$ & $>0.60$       & $>0.40$       \nl
      &  $+311$ & $0.55\pm0.53$ &               & unconstrained & unconstrained \nl
      &  $+336$ & $0.89\pm0.17$ &               & $>0.52$       & $>0.14$       \nl
      &  $+365$ & $0.67\pm0.09$ &               & $>0.29$       & unconstrained \nl
\enddata
\tablenotetext{a}{This table gives the weighted-mean coverage
fractions of detected NALs, averaged over small velocity bins of width
20--40~{\kms}.  Each bin is labelled by its mean velocity.}
\tablenotetext{b}{The reported values of $\cf$ and $\langle v\rangle$
are weighted averages over the listed velocity range. Weights were
assigned according to the variance in each resolution element's
computed coverage fraction.}
\tablenotetext{c} {The range of allowed values of $\cc$ and $\celr$
based on the measured values of $\cf$ and $W$. The limits are
determined by applying equation~(5) and using the $1\sigma$
uncertainties on the measured quantities. They correspond to the
corners of the extreme boxes of allowed solutions similar to those
illustrated in Figure~\ref{fig:illust}.  As such they cannot hold
independently of each other because the solution must also satisfy
equation~(5).}
\label{tab:covfac2} 
\end{deluxetable}

                
                
\clearpage

\begin{deluxetable}{cccccc}
\tablewidth{0pc}
\tablecolumns{6}
\tablenum{4}
\tablecaption{Coverage Fractions from Literature}
\tablehead
{
\colhead{Ion} &  
\colhead{$\langle v\rangle$} &  
\colhead{W} &
\colhead{$\cf$} & 
\colhead{$\cc$\tablenotemark{a}} & 
\colhead{$\celr$\tablenotemark{a}} \nl
\colhead{} &
\colhead{(\kms)} &
\colhead{} &
\colhead{} &
\colhead{} &
\colhead{} 
}
\startdata
\cutinhead{Q~$0123+237$ \hskip 1cm (radio loud; \cite{bs97})}
\CIV  & $ +190$  & 1.0  & 0.63$^{+0.07}_{-0.04}$ & $>$0.26 & $>$0.26 \nl
\CIV  & $ +480$  & 1.0  & 0.94$^{+0.04}_{-0.03}$ & $>$0.88 & $>$0.88 \nl
\NV   & $ +190$  & 0.3  & 0.79$^{+0.21}_{-0.17}$ & $>$0.73 & $>$0.09 \nl
\NV   & $ +480$  & 0.3  & 0.98$^{+0.02}_{-0.02}$ & $>$0.98 & $>$0.93 \nl
\SiIV & $ +480$  & 0.2  & 0.93$^{+0.07}_{-0.05}$ & $>$0.92 & $>$0.57 \nl
\Lya  & $ +190$  & 0.9  & $>0.91$                & $>$0.80 & $>$0.80 \nl
\Lya  & $ +480$  & 0.9  & $>0.85$                & $>$0.71 & $>$0.68 \nl
\cutinhead{Q~$0150-203$ \hskip 1cm (radio quiet; \cite{ham97a})}
\CIV  & $-1520$  & 1.2  & 0.46$^{+0.1}_{-0.0}$   & unconstrained & $<$0.80 \nl
\CIV  & $-1400$  & 1.1  & 0.44$^{+0.1}_{-0.0}$   & $<$0.97 & $<$0.80 \nl
\NV   & $-1700$  & 0.8  & 0.32$^{+0.3}_{-0.0}$   & $<$0.58 & $<$0.72 \nl
\NV   & $-1520$  & 0.8  & 0.63$^{+0.2}_{-0.1}$   & $>$0.33 & $>$0.17 \nl
\NV   & $-1400$  & 0.8  & 1.00$^{+0.0}_{-0.1}$   & 1.00      & 1.00 \nl
\NV   & $-1300$  & 0.8  & 1.00$^{+0.0}_{-0.1}$   & 1.00      & 1.00 \nl
\cutinhead{Q~$2343+125$ \hskip 1cm (radio quiet; \cite{ham97b})}
\CIV  & $-24,000$\tablenotemark{b} & 0.1 & 0.045$^{+0.03}_{-0.00}$ & $<$0.50 & $<$0.50 \nl
\CIV  & $-24,000$\tablenotemark{c} & 0.1 & 0.19$^{+0.04}_{-0.02}$  & 0.11--0.21 & unconstrained \nl
\enddata
\tablenotetext{a}{The limits on $\cc$ and $\celr$ have the same
significance as in Table~\ref{tab:covfac2}.}
\tablenotetext{b}{As observed in 1994 September.}
\tablenotetext{c}{As observed in 1995 October.}
\label{tab:covfac}
\end{deluxetable}

\clearpage

\clearpage

\clearpage

\end{document}